\documentclass[a4paper]{jpconf}
\usepackage{graphicx}
\usepackage{epstopdf}
\usepackage{amsmath}
\usepackage{cite}
\newcommand{\pT}{p_{\rm T}}
\newcommand{\Raa}{R_{\rm AA}}

\begin{document}
\title{D-meson nuclear modification factor and v$_2$ in Pb-Pb collisions at the LHC}

\author{Elena Bruna, for the ALICE Collaboration}

\address{Istituto Nazionale di Fisica Nucleare, Via P. Giuria 1, 10125 Torino}

\ead{bruna@to.infn.it}

\begin{abstract}
%The remarkable heavy-flavour results from RHIC and LHC show that heavy quarks are affected by the strongly coupled medium created in heavy-ion collisions at high energies. A way to characterize the properties of such hot and dense medium is to quantify to what extent the medium influences the propagation of the particles through it, and how that is different for different parton species.
%Heavy quarks are produced in the early stages of a heavy-ion collision and, given their large mass, their abundance is not expected to change throughout the evolution of the system. Therefore, they behave as self-generated probes that carry information of the medium by losing energy via subsequent interactions.
%ALICE provides high precision tracking, particle identification and excellent vertexing capabilities, which allow measurements of heavy-flavoured particles in a wide momentum range. 
%The remarkable results from RHIC and LHC show that a strongly coupled medium is created in heavy-ion collisions at high energies. A way to characterize the properties of such hot and dense medium is via the heavy quarks, which are initially produced probes carrying information of the medium by losing energy via subsequent interactions.
%ALICE provides high precision tracking, particle identification and excellent vertexing capabilities, which allow measurements of heavy-flavoured particles in a wide momentum range. 

We present the ALICE results on open heavy flavour, focusing on the exclusive reconstruction of charmed mesons via displaced decay topologies.
These measurements benefit from the large Pb-Pb statistics collected in 2011.
The results on the nuclear modification factor $\Raa$ for $D$ mesons indicate a suppression of their yield in central collisions relative to binary-scaled pp collisions in a large momentum range. The comparison to the $\Raa$ of non-prompt J/$\psi$ (measured with CMS) indicates a difference in the suppression of charm and beauty, as expected according to the predicted mass hierarchy in energy loss models.
 The measurement of the azimuthal anisotropy of charmed mesons is also discussed. The observed positive second Fourier harmonic $v_2$ for transverse momentum $2<\pT<6$ GeV/$c$ in semi-peripheral events is a hint for collective motion of charm quarks. The results discussed above are also compared to theoretical models. 
\end{abstract}

\section{Introduction}

A quantitative picture of the medium created in high-energy Pb-Pb collisions can be obtained by measuring its properties with observables that give access to the energy-loss mechanism for different parton species and to the transport properties of the medium.
Theoretical models of energy loss predict a hierarchical dependence on the colour charge and mass of the parton propagating through the medium. A larger energy loss is expected for gluons, together with a suppression of gluon radiation at small angles for partons with larger mass (``dead-cone" effect~\cite{deadcone}). 
It is therefore interesting to compare medium effects (i.e. the path-length, colour-charge and mass dependence of the energy loss, as well as the collective motion) on heavy quarks versus light quarks and gluons.
Such challenging goals require precise heavy-flavour measurements in Pb-Pb collisions and a clear understanding of the reference systems provided by pp and p-Pb collisions. The latter provides the control experiment needed to disentangle effects from the cold matter in absence of a hot and dense medium.

\section{Heavy-flavour measurements in ALICE}
%ALICE~\cite{alice} is equipped with low magnetic field and high-precision silicon and gas detectors with low-material budget that provide primary and secondary vertex reconstruction, as well as tracking and particle identification down to transverse momenta of 100 MeV/$c$.  In addition, the VZERO detector is used for trigger and centrality measurements.
%that allow to m and particle identification down to transverse momenta of 100 MeV/$c$. These features are essential ingredients to perform open-heavy flavor analyses over a broad transverse momentum range. %The ALICE detectors used in these analyses are the Inner Tracking System (ITS) for tracking as well as primary and secondary vertex reconstruction, the Time Projection Chamber (TPC) for tracking and particle identification, the Time Of Flight (TOF) for particle identification and the VZERO for trigger and centrality measurements. For more details refer to~\cite{alice}.
The ALICE~\cite{alice} detectors used in these analyses are the Inner Tracking System (ITS), the Time Projection Chamber (TPC) and the Time Of Flight (TOF) detector. They provide tracking, primary and secondary vertex reconstruction and particle identification, like $K/\pi$ separation up to transverse momentum $\pT\sim~2.5$ GeV/$c$.  These features are essential ingredients to perform open-heavy flavour analyses over a broad transverse momentum range. In addition, the VZERO detector is used for triggering on collisions and for centrality measurements. %For more details refer to~\cite{alice}.
Charmed mesons are reconstructed in ALICE through their hadronic decays: $D^0 \rightarrow K^-\pi^+$, $D^+ \rightarrow K^-\pi^+ \pi^+$, $D^{*+} \rightarrow D^0 \pi^+$ and $D_s^+ \rightarrow \phi (\rightarrow K^-K^+)\pi^+$. The analysis strategy exploits the particle identification and the detector capability to reconstruct secondary vertices displaced by a few hundred $\mu$m from the primary vertex. Topological cuts are applied on the reconstructed secondary vertices and the signals are extracted via fits to the invariant mass distributions.
%The results presented here are the prompt D-meson nuclear modification factor R$_{AA}$\footnote{R$_{AA}$ is defined as the yield in Pb-Pb collision divided by the cross-section in pp collisions scaled by the average nuclear overlap function T$_{AA}$ in a given Pb-Pb centrality range.} as a function of $\pT$ and collision centrality and the elliptic flow $v_2$\footnote{The elliptic flow is related to the azimuthal anisotropy of  particle emission in case of non-central A-A collisions. If the medium is characterized by a large collectivity of its particles, the typical almond-shaped initial geometry of a non-central collision is  transformed into an anisotropy in momentum space which is experimentally quantified by a non-zero second harmonic coefficient $v_2$ of the Fourier azimuthal expansion. At higher momentum an interplay of collective motion with a path-length dependence of the in-medium energy loss is also expected given the different paths traveled by a parton propagating through the medium. 
%} as a function of $\pT$ in semi-peripheral collisions.

In these proceedings we report on the prompt $D$-meson nuclear modification factor $\Raa$, defined as the yield in Pb-Pb collisions divided by the cross-section in pp collisions scaled by the average nuclear overlap function in a given Pb-Pb centrality range, as a function of $\pT$ and collision centrality. The prompt $D$-meson elliptic flow $v_2$, defined as the second harmonic coefficient of the Fourier azimuthal expansion, is also reported as a function of $\pT$ in semi-peripheral collisions.

In the $\Raa$ analysis, the prompt $D$-meson yield was obtained after subtracting the contribution of $D$ mesons from beauty hadron decays based on perturbative QCD calculations (FONLL,~\cite{fonll}). The $\Raa$ of non-prompt $D$ mesons was assumed to be  $\Raa^{\rm {feed-down}}=2\Raa^{\rm{prompt}}$(on the basis of the CMS results of non-prompt $J/\psi$, instead of  $\Raa^{\rm{feed-down}}=\Raa^{\rm{prompt}}$ as done in the previous analysis~\cite{Raa2010}) and the systematic uncertainty was assessed by varying its value between 1 and 3 times the prompt-$D$ R$_{AA}$. %the corresponding systematic uncertainty was estimated to be $\sim 6-10 \%$.
Due to the lack of statistics of the pp data at $\sqrt s=2.76$ TeV, the pp reference needed for the R$_{AA}$ analysis was determined by scaling the $D$-meson cross section measured at $\sqrt s=7$ TeV to  $\sqrt s=2.76$ TeV using FONLL calculations~\cite{scaling}. The scaling uncertainty varies from $\sim 20 \%$ to $\sim 5 \%$ from low to high $\pT$.
The contribution of $D$ mesons from beauty  decays was also subtracted from the $v_2$ measurement with an assumption on the $\Raa$ (as above) and $v_2$ of non-prompt $D$ mesons ($0 \leq v_2^{\rm{feed-down}} \leq v_2^{\rm{prompt}}$~\cite{Dflow}). %This results in a systematic uncertainty up to $\sim 45 \%$~\cite{Dflow}.
These analyses were performed on the data from the Pb-Pb runs at $\sqrt s_{\rm{NN}}=2.76$ TeV collected in 2010 (minimum bias trigger, $\mathcal{L} =2.12~\mu$b$^{-1}$) and 2011 ($\mathcal{L} =28$ and $6~\mu$b$^{-1}$ for 0-7.5\% and 10-50\% respectively).

\section{Results}

\begin{figure}
\centering
\resizebox{0.39\textwidth}{!}{  \includegraphics{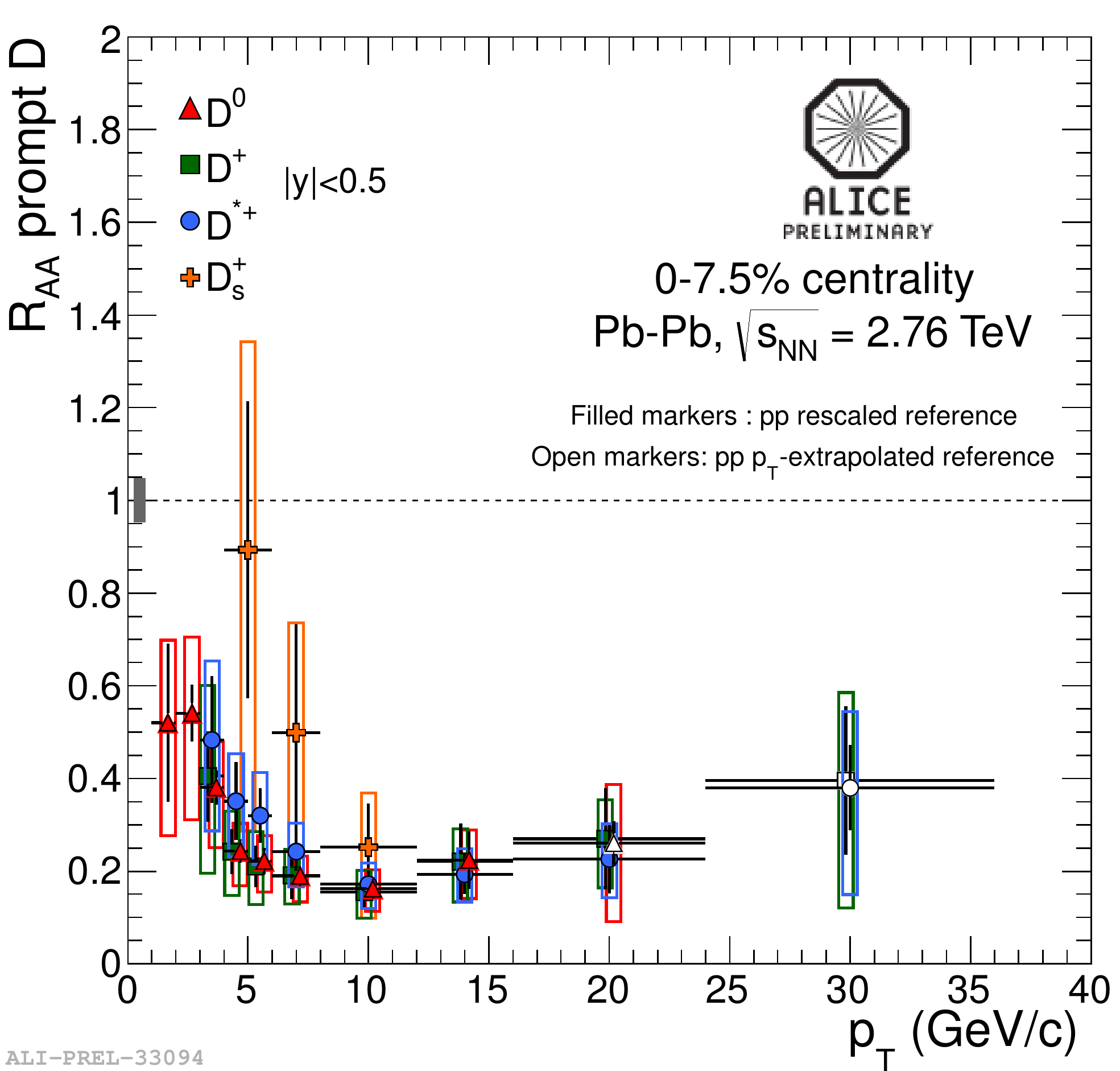}}
\resizebox{0.38\textwidth}{!}{  \includegraphics{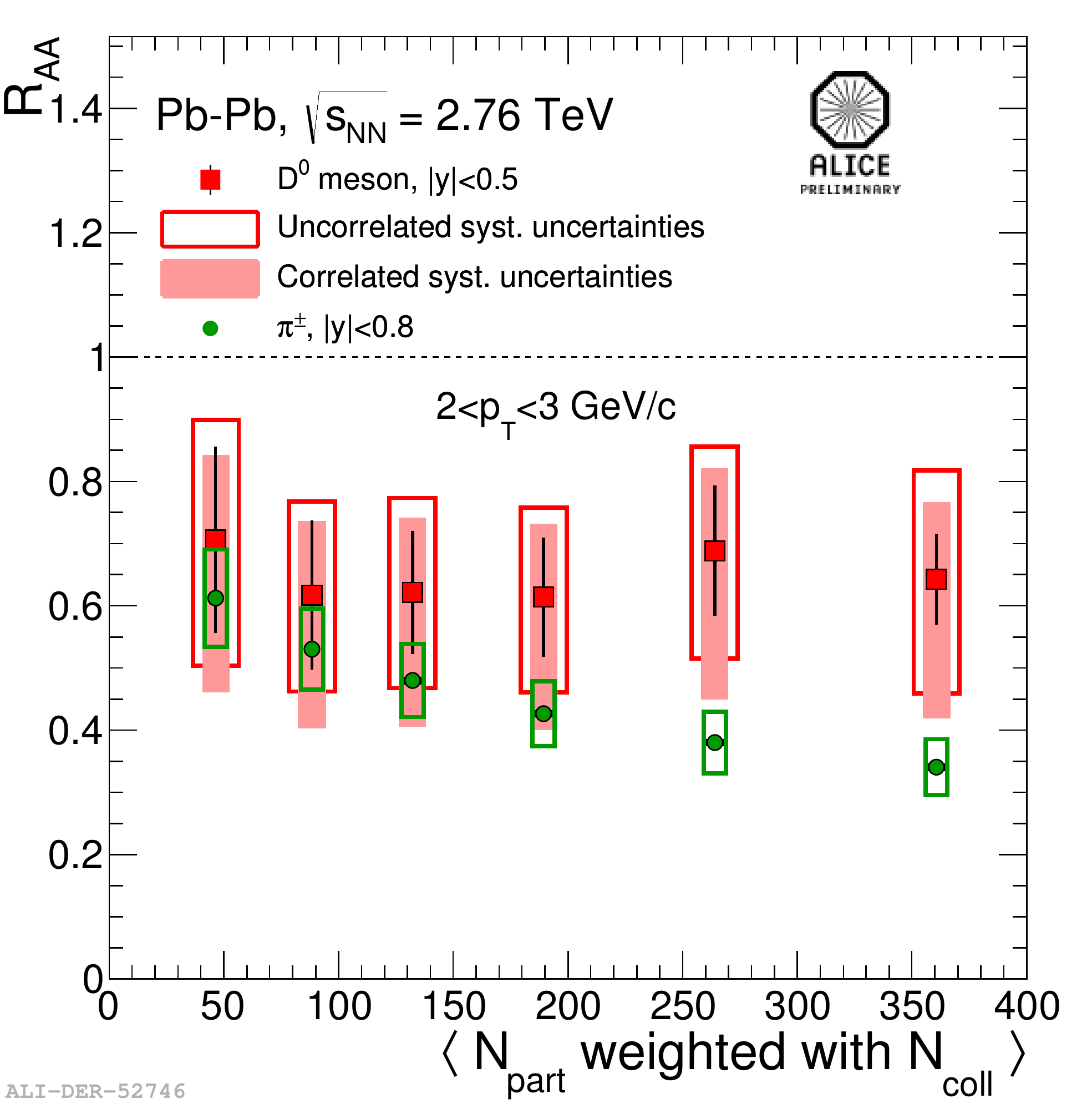}}
\caption{Left: $\Raa$ as a function of $\pT$ for prompt $D^0$, $D^+$, $D^{*+}$ and $D_s^+$ for the $0-7.5\%$ most central Pb-Pb collisions at $\sqrt s_{\rm NN}=2.76$ TeV~\cite{zaidaQM}. Right: $\Raa$ as a function of centrality for prompt $D^0$ mesons in the transverse momentum range  $2<\pT<3$ GeV/$c$, compared to pions.}
\label{fig:Raa}
\end {figure}

\begin{figure}
\centering
\resizebox{0.35\textwidth}{!}{  \includegraphics{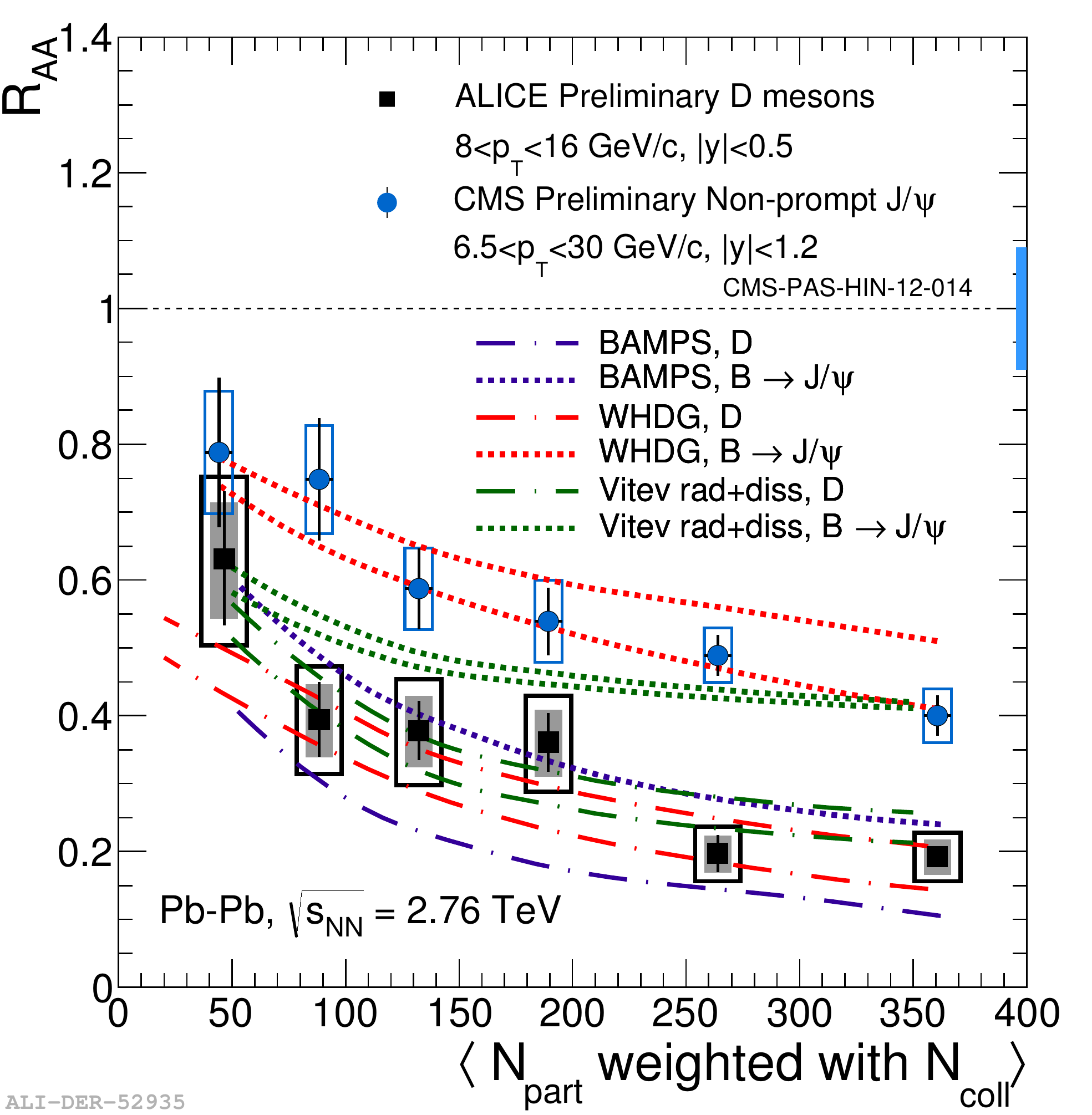}}
\resizebox{0.45\textwidth}{!}{  \includegraphics{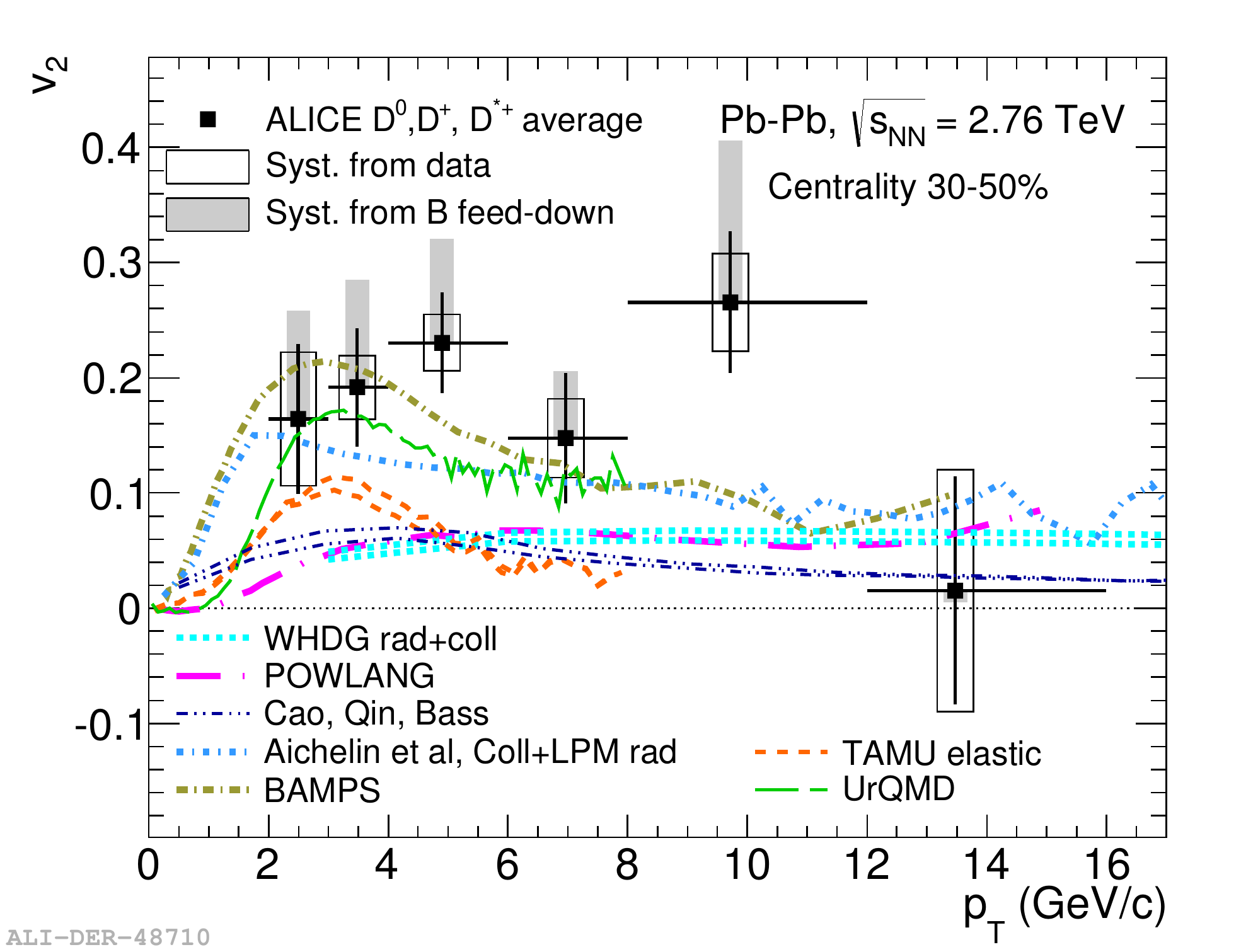}}
\caption{Left: $\Raa$ as a function of centrality for prompt $D$ mesons (average of $D^0$, $D^+$ and $D^{*+}$) in the transverse momentum range  $8<\pT<16$ GeV/$c$, compared to non-prompt J/$\psi$ measured with CMS with $6.5<\pT<30$ GeV/$c$. Results from theoretical calculations are superimposed (see text). Right: $v_2$ for prompt $D$ mesons in the 30-50\% centrality class as a function of $\pT$~\cite{Dflow} superimposed to theoretical results.}
\label{fig:RaaB}
\end {figure}

The $\Raa$ of prompt $D$ mesons was measured with ALICE in the $0-7.5\%$ centrality class using the data sample collected in 2011, which extends the measurements to a wider transverse momentum range ($2<\pT<36$ GeV/$c$) compared to the published results from the 2010 data~\cite{Raa2010}.
As shown in Fig.~\ref{fig:Raa} (left), the $\Raa$ values for  $D^0$, $D^+$ and $D^{*+}$ agree within the uncertainties and show a strong suppression (factor of $4-5$ for $5<\pT<16$ GeV/$c$) of the $D$-meson yields in Pb-Pb collisions relative to pp collisions. The first measurement of the $D_s^+$ meson in Pb-Pb collisions is also shown. A suppression of the  $D_s^+$ is observed for $8<\pT<12$ GeV/$c$, in agreement within the uncertainties with the non-strange charmed mesons $\Raa$ in this $\pT$ range. %However, at lower $\pT$, the yield of $D_s$ mesons could be suppressed less and more data is needed to draw a conclusion. Theoretical models suggest an enhancement of low-$\pTpart$ $D_s$ mesons  relative to non-strange D mesons due to c-quark recombination with the enhanced strange quarks in the medium~\cite{DsTheory}.
The $D_s^+$-meson yield could be less suppressed at lower $\pT$ because of the predicted c-quark recombination with the enhanced strange quarks in the medium~\cite{DsTheory}, however a more precise measurement is needed to draw a conclusion. 
The $\Raa$  as a function of the number of participant nucleons weighted by the number of binary collisions is shown in Fig.~\ref{fig:Raa} (right) for $D^0$ mesons with $2<\pT<3$~{\rm GeV}/$c$ and compared to pions measured in the same centrality and $\pT$  intervals. The main correlated systematic uncertainties for the $D$-meson yields %those correlated in the different centrality bins (i.e. % 
come from the pp cross section and beauty feed down. The uncorrelated systematic uncertainties are mainly due to the $D$-meson yield extraction and cut variation. 
%The same suppression pattern of D mesons is observed in different centralities in the $2<\pT<3$ GeV/$c$ range and a similar suppression as for the pions is observed  within uncertainties. A hint of a difference in central collisions is suggested but more statistics is needed to draw conclusion on the expected difference between D and pion suppression due to the mass and colour charge dependence of the energy loss.
The suppression of $D$ mesons with $2<\pT<3$ GeV/$c$ is almost independent of centrality, while the pion $\Raa$ decreases with increasing centrality. The magnitude of the suppression is similar for $D$ mesons and pions. A hint of a difference in central collisions is suggested but a larger dataset is needed to draw conclusion on the expected difference between $D$ mesons and pion suppression due to the mass and colour charge dependence of the energy loss.
In the higher $\pT$ range the $D$-meson suppression is no longer flat, rather it increases from peripheral to central collisions (Fig.~\ref{fig:RaaB}, left).
%The beauty frontier in Pb-Pb collisions has been explored by CMS, in particular with the measurement of the $R_{AA}$ of secondary J/$\psi$ from beauty decays.

The $\Raa$ of non-prompt  J/$\psi$ ($6.5 < \pT<30$ GeV/$c$ and $|y|<1.2$) measured with CMS is reported in Fig.~\ref{fig:RaaB} (left) as a function of the number of participants weighted by the number of binary collisions~\cite{CMSnonprompt}. The results are shown together with the $D$-meson $\Raa$ from ALICE  ($8 < \pT<16$ GeV/$c$ and $|y|<0.5$). The selected $\pT$ ranges correspond to similar kinematical ranges for the parent b and c quarks, but the measurements are performed in different rapidity ranges. An indication of a difference in the suppression of charm and beauty can be observed in central collisions, consistent with the mass hierarchy expected from various energy-loss models, such as those reported in the figure~\cite{BAMPS,WHDG,vitev}. While the model~\cite{WHDG} based on collisional and radiative energy loss in an anisotropic medium agrees with both prompt $D$ mesons and non-prompt J/$\psi$, the others seem to underestimate both (i.e.~\cite{BAMPS}, based on collisional energy loss in an expanding medium) or to underestimate the suppression of non-prompt J/$\psi$ (i.e.~\cite{vitev}, based on radiative energy loss and $D$-meson in-medium formation and dissociation).
%The other experimental observable used to characterize the hot and dense medium is the ``elliptic flow", related to the azimuthal anisotropy of  particle emission in case of non-central A-A collisions. If the medium is characterized by a large collectivity of its particles, the typical almond-shaped initial geometry of a non-central collision is  transformed into an anisotropy in momentum space which is experimentally quantified by a non-zero second harmonic coefficient $v_2$ of the Fourier azimuthal expansion.

The $v_2$ for prompt $D$ mesons, reported in Fig.~\ref{fig:RaaB} (right), was measured as a function of $\pT$ in the range  $2 <\pT<16$ GeV/$c$ in the 30-50\% centrality class~\cite{Dflow}.
We observe a positive $v_2$ for prompt $D$ mesons  in the transverse momentum range $2<\pT<6$ GeV/$c$, which is comparable to that of charged particles (not shown).
This result suggests a collective motion also for heavy quarks at low $\pT$.  At higher momentum a positive $v_2$ is
expected due to the path-length dependence of the in-medium energy loss,
given the different paths traveled by a parton propagating through the medium. 
The measurements of the prompt $D$-meson $v_2$ are also compared to various energy-loss models~\cite{BAMPS,WHDG, POWLANG,Aichelin,TAMU,UrQMD}.
While several theoretical models based on in-medium parton energy loss reproduce reasonably well the $\Raa$ of prompt $D$ mesons as a function of $\pT$ (not shown, see~\cite{Raa2010}), they are challenged by simultaneously reproducing results from prompt $D$-meson $\Raa$  and $v_2$.

\section{Conclusions}
 The  suppression of prompt $D$ mesons observed in the momentum range $5<\pT<15$ GeV/$c$ in central Pb-Pb collisions at $\sqrt s_{\rm NN}=2.76$ TeV is evidence for a strong modification of charm production in Pb-Pb compared to pp collisions due to energy loss. Results for the $D_s^+$ meson also show a suppression in $8<\pT<12$ GeV/$c$ but suffer from statistics limitations that do not allow us to draw conclusions on a possible enhancement  at low $\pT$ due to a coalescence of charm quarks with strange quarks.
 The measurements of prompt $D$-meson $\Raa$ as a function of centrality indicate that the suppression tends to be constant with centrality in the lowest $\pT$ range, $2<\pT<3$ GeV/$c$, while it increases with centrality at intermediate/high $\pT$. The observed difference in the suppression of $D$ mesons and non-prompt J/$\psi$ from B-meson decays at high $\pT$ in central collisions suggests a mass dependence in the energy loss of heavy quarks.
 The positive $v_2$ for prompt $D$ mesons  in the momentum range $2<\pT<6$ GeV/$c$ in semi-peripheral Pb-Pb collisions suggests that the interactions
with the medium constituents transfer to charm quarks information on the
azimuthal anisotropy of the system. 
 Theoretical models of in-medium energy loss reproduce reasonably well the measured prompt $D$-meson $\Raa$. Nevertheless,   a simultaneous description of both $\Raa$ and $v_2$ remains a challenge for models, indicating that the energy-loss mechanisms
and the participation of heavy quarks in the collective behavior of the Quark-Gluon Plasma
are not yet fully understood today.
%it is still challenging to have a simultaneous description of both $R_{AA}$ and $v_2$, making this still an interesting open point to understand the energy loss mechanisms and collective behavior of heavy quarks.


\section*{References}
\begin{thebibliography}{9}

\bibitem{deadcone}
  Y.~L.~Dokshitzer and D.~E.~Kharzeev,
  %``Heavy quark colorimetry of QCD matter,''
  Phys.\ Lett.\ B {\bf 519} (2001) 199
  [hep-ph/0106202].

\bibitem{alice}
  K.~Aamodt {\it et al.}  [ALICE Collaboration],
  %``The ALICE experiment at the CERN LHC,''
  JINST {\bf 3} (2008) S08002.
  %%CITATION = JINST,3,S08002;%%
  %501 citations counted in INSPIRE as of 10 Oct 2013
  
\bibitem{fonll}
 % M.~Cacciari, S.~Frixione, N.~Houdeau, M.~L.~Mangano, P.~Nason and G.~Ridolfi,
  M.~Cacciari {\it et al.},
  %``Theoretical predictions for charm and bottom production at the LHC,''
  JHEP {\bf 1210} (2012) 137
  [arXiv:1205.6344].
  %%CITATION = ARXIV:1205.6344;%%
  %42 citations counted in INSPIRE as of 10 Oct 2013

\bibitem{Raa2010}
  B.~Abelev {\it et al.}  [ALICE Collaboration],
  %``Suppression of high transverse momentum D mesons in central Pb-Pb collisions at $\sqrt{s_{NN}}=2.76$ TeV,''
  JHEP {\bf 1209} (2012) 112
  [arXiv:1203.2160].
  %%CITATION = ARXIV:1203.2160;%%
  %88 citations counted in INSPIRE as of 10 Oct 2013
  
\bibitem{scaling}
  R.~Averbeck {\it et al.},
  %``Reference Heavy Flavour Cross Sections in pp Collisions at \surd s = 2.76 TeV, using a pQCD-Driven \surd s-Scaling of ALICE Measurements at \surd s = 7 TeV,''
  arXiv:1107.3243.
  %%CITATION = ARXIV:1107.3243;%%
  %24 citations counted in INSPIRE as of 10 Oct 2013

\bibitem{Dflow}
  B.~Abelev {\it et al.}  [ALICE Collaboration],
  %``D meson elliptic flow in non-central Pb-Pb collisions at sqrts(s_NN) = 2.76TeV,''
  Phys.\  Rev.\  Lett.\  {\bf 111} (2013) 102301
  [arXiv:1305.2707].
  %%CITATION = ARXIV:1305.2707;%%
  %6 citations counted in INSPIRE as of 10 Oct 2013

\bibitem{zaidaQM}
  Z.~Conesa del Valle [ALICE Collaboration],
  %``Heavy-flavor suppression and azimuthal anisotropy in Pb-Pb collisions at $\sqrt(s_{NN}) = 2.76$ TeV with the ALICE detector,''
  Nucl.\ Phys.\ A {\bf 904-905} (2013) 178c
  [arXiv:1212.0385].
  %%CITATION = ARXIV:1212.0385;%%
  %7 citations counted in INSPIRE as of 10 Oct 2013
  
\bibitem{DsTheory}
  I.~Kuznetsova {\it et al.},
  %``Heavy flavor hadrons in statistical hadronization of strangeness-rich QGP,''
  Eur.\ Phys.\ J.\ C {\bf 51} (2007) 113.
   M.~He {\it et al.},
  %``$\mathbf{D_s}$-Meson as Quantitative Probe of Diffusion and Hadronization in Nuclear Collisions,''
  Phys.\ Rev.\ Lett.\  {\bf 110} (2013) 112301.
 A.~Andronic {\it et al.},
  %``Charmonium and open charm production in nuclear collisions at SPS/FAIR energies and the possible influence of a hot hadronic medium,''
  Phys.\ Lett.\ B {\bf 659} (2008) 149.
  %%CITATION = ARXIV:0708.1488;%%
  %33 citations counted in INSPIRE as of 10 Oct 2013
  %%CITATION = HEP-PH/0607203;%%
  %34 citations counted in INSPIRE as of 10 Oct 2013

\bibitem{CMSnonprompt}
  S.~Chatrchyan {\it et al.}  [CMS Collaboration],
  %``Suppression of non-prompt $J/\psi$, prompt $J/\psi$, and Y(1S) in PbPb collisions at $\sqrt{s_{NN}}=2.76$ TeV,''
  JHEP {\bf 1205} (2012) 063
  [arXiv:1201.5069], CMS-PAS-HIN-12-014.
  %%CITATION = ARXIV:1201.5069;%%
  %104 citations counted in INSPIRE as of 10 Oct 2013

\bibitem{BAMPS}
  O.~Fochler, J.~Uphoff, Z.~Xu and C.~Greiner,
  %``Jet quenching and elliptic flow at RHIC and LHC within a pQCD-based partonic transport model,''
  J.\ Phys.\ G {\bf 38} (2011) 124152
  [arXiv:1107.0130].
  %%CITATION = ARXIV:1107.0130;%%
  %16 citations counted in INSPIRE as of 10 Oct 2013

\bibitem{WHDG}
  W.~A.~Horowitz and M.~Gyulassy,
  %``Quenching and Tomography from RHIC to LHC,''
  J.\ Phys.\ G {\bf 38} (2011) 124114
  [arXiv:1107.2136].
  %%CITATION = ARXIV:1107.2136;%%
  %27 citations counted in INSPIRE as of 10 Oct 2013
  
\bibitem{vitev}
  I.~Vitev {\it et al.},
  %``Light-cone wave function approach to open heavy flavor dynamics in QCD matter,''
  Phys.\ Rev.\ C {\bf 80} (2009) 054902,
  %Phys.\ Lett.\ B {\bf 713} (2012) 224. %jet,di-jet

\bibitem{POWLANG}
  W.~M.~Alberico {\it et al.},
  %``Heavy-flavour spectra in high energy nucleus-nucleus collisions,''
  Eur.\ Phys.\ J.\ C {\bf 71} (2011) 1666
  [arXiv:1101.6008].
  %%CITATION = ARXIV:1101.6008;%%
  %44 citations counted in INSPIRE as of 10 Oct 2013

\bibitem{Aichelin}
  P.~B.~Gossiaux {\it et al.},
  %``Tomography of a quark gluon plasma at RHIC and LHC energies,''
  Phys.\ Rev.\ C {\bf 79} (2009) 044906,
  %``Competition of Heavy Quark Radiative and Collisional Energy Loss in Deconfined Matter,''
  J.\ Phys.\ G {\bf 37} (2010) 094019.

\bibitem{TAMU} Rapp, He {\it et al.}, Phys. Rev. C 86 (2012) 014903.
\bibitem{TAMU}
  M.~He, R.~J.~Fries and R.~Rapp,
  %``Heavy-Quark Diffusion and Hadronization in Quark-Gluon Plasma,''
  Phys.\ Rev.\ C {\bf 86} (2012) 014903
  [arXiv:1106.6006].
  %%CITATION = ARXIV:1106.6006;%%
  %29 citations counted in INSPIRE as of 10 Oct 2013

\bibitem{UrQMD}
  T.~Lang {\it et al.},
  %``Heavy quark transport in heavy ion collisions at RHIC and LHC within the UrQMD transport model,''
  arXiv:1211.6912,  arXiv:1212.0696.
  %%CITATION = ARXIV:1211.6912;%%
  %9 citations counted in INSPIRE as of 10 Oct 2013

\end{thebibliography}
\end{document}